\documentclass[%
 reprint,
superscriptaddress,
 amsmath,amssymb,
 aps,
prl,
]{revtex4-2}
\usepackage[resetlabels,labeled]{multibib}
\usepackage{hyperref}
\hypersetup{
    colorlinks=true,
    citecolor=blue,
    linkcolor=blue,
    urlcolor=blue
    }

\usepackage{soul}
\usepackage{booktabs} %use for tables
\usepackage{graphicx}% Include figure files
\usepackage{dcolumn}% Align table columns on decimal point
\usepackage{bm}% bold math
\usepackage{mathptmx}
\usepackage[dvipsnames]{xcolor} % Needed for \hgh command. See this link for details -> https://en.wikibooks.org/wiki/LaTeX/Colors#The_68_standard_colors_known_to_dvips
\usepackage{float} % Otherwise, you cannot use the [H] option in figures [H] forces the figure to be displayed exactly where the code is written

\usepackage{orcidlink}

\usepackage[normalem]{ulem}

\usepackage[normalem]{ulem}

\newcommand{\sm}{\ensuremath{\mathbf{S}}}
\begin{document}

\preprint{APS/123-QED}

\title{Efficient motion of 90$^{\circ}$ domain walls in Mn$_2$Au via pure optical torques} % Force line breaks with \\

%\Oks{I think that the title "Efficient motion of 90$^{\circ}$ domain walls in antiferromagnet Mn$_2$Au by purely optical torque" is more correct that your title. There are many other ways to move DWs by laser, for example, by thermal gradient created by laser. }

\author{Paul-Iulian Gavriloaea\orcidlink{0000-0001-7519-4792}}
\affiliation{Instituto de Ciencia de Materiales de Madrid, CSIC, Cantoblanco, 28049 Madrid, Spain}
\author{Jackson L. Ross\orcidlink{0000-0001-5417-8216}}
\affiliation{School of Physics, Engineering and Technology, University of York, YO10 5DD, York, United Kingdom}
\author{Frank Freimuth\orcidlink{0000-0001-6193-5991}}
\affiliation{Institute of Physics, Johannes Gutenberg University Mainz, 55099 Mainz, Germany}
\affiliation{Peter Grünberg Institut and Institute for Advanced Simulation, Forschungszentrum Jülich and JARA, 52425 Jülich, Germany}
\author{Yuriy Mokrousov\orcidlink{0000-0003-1072-2421}}
\affiliation{Peter Grünberg Institut and Institute for Advanced Simulation, Forschungszentrum Jülich and JARA, 52425 Jülich, Germany}
\affiliation{Institute of Physics, Johannes Gutenberg University Mainz, 55099 Mainz, Germany}
\author{Richard~F.~L.~Evans\orcidlink{0000-0002-2378-8203}}
\affiliation{School of Physics, Engineering and Technology, University of York, YO10 5DD, York, United Kingdom}
\author{Roy Chantrell \orcidlink{0000-0001-5410-5615}}
\affiliation{School of Physics, Engineering and Technology, University of York, YO10 5DD, York, United Kingdom}
\author{Oksana Chubykalo-Fesenko\orcidlink{0000-0002-4081-1831}}
\affiliation{Instituto de Ciencia de Materiales de Madrid, CSIC, Cantoblanco, 28049 Madrid, Spain}
\author{Rubén M. Otxoa\orcidlink{0000-0003-1534-4159}}
\affiliation{Hitachi Cambridge Laboratory, CB3 OHE Cambridge, United Kingdom}
\affiliation{Donostia International Physics Center, 20018 Donostia San Sebastian, Spain}

\begin{abstract}
%\Oks{The abstract in PRL should be short, it should be like 3 sentences. You should not refer that your results follow other results because this decreases the novelty. I have commented the previous abstract and shortened it. See how you like it. Also in PRL the abstract should start with the introductory sentence}

Discovering alternative ways to drive  domain wall (DW)  dynamics is crucial for advancing spintronic applications.
%We report on the theoretical demonstration of laser-induced
 %90$^{\circ}$ domain-wall (DW) dynamics in an antiferromagnetic (AFM) system. Previous \textit{ab-initio} results predict a linearly or circularly polarised laser excitation at optical frequencies can induce a substantial torque on the N\'eel order parameter of the Mn$_2$Au crystal. This torque arises via a staggered opto-magnetic field
 %\Jackson{optical field; optical field is perhaps misleading, as it is induced from the optics but not the field of the light} !! I believe now its clear the opto-magnetic coupling leads to this torque. 
 % of tens of $mT$ \Rub{Maybe no need to specify order of magnitude and save some space in the abstract} !! removed this
% which couples with the distinct Mn spins in a similar picture to the relativistic N\'eel Spin Orbit Torque. 
 Here we demonstrate via atomistic spin dynamics simulations that 
 %the theoretically predicted
 optical torques can efficiently drive 90$^{\circ}$ DWs in the Mn$_2$Au antiferromagnet  but
 %at \sout{superluminal} \Oks{not a good wording because in reality this is just an analogy, supermagnonic?} velocities, but %interestingly 
 their spatial symmetry forbids the motion of 180$^{\circ}$  walls. 
 In the steady-state regime, the kinematics display special relativity signatures
 %with the increase of the laser pulse intensity: the propagation velocity saturates up to a given magnonic limit of 46 km/s equivalent to an effective “speed of light”, whilst the wall-width exhibits a decrease in length, well described by a Lorentz
%contraction factor. 
%Interestingly, near the magnonic limit, 
accessed for low laser intensities. 
%as low as $0.3$ GW/cm$^2$. 
At  velocities higher than the magnonic limit, the  DW enters a proliferation regime in which part of its relativistic energy is invested into the nucleation of novel magnetic textures. 
  %We believe 
  Our investigation contributes towards the fundamental understanding of opto-magnetic effects, supporting the development of next generation, all-optically controlled antiferromagnetic spintronics.

\end{abstract}

\maketitle

%\tableofcontents

%\section{\label{sec:level1}Introduction}
%\textbf{I think this can be shortened}\\
Antiferromagnets (AFMs) are envisioned as promising active elements for the next-generation of spintronic devices \cite{Jungwirth2016AntiferromagneticSpintronics,Baltz2018AntiferromagneticSpintronics}.
%,Jungwirth2018TheSpintronics
To this end, a rapid and efficient manipulation of magnetic domain walls (DWs) is desired, aiming towards the realisation of race-track memories \cite{Parkin2008MagneticMemory,Yang2015Domain-wallAntiferromagnets}, logic architectures \cite{Allwood2005MagneticLogic,Lan2017AntiferromagneticRetarder, Hedrich2021NanoscaleWalls}  or unconventional computing protocols \cite{Zhang2020Antiferromagnet-BasedCharges,Ababei2021NeuromorphicWall,Otxoa2022TailoringComputing}. In ferromagnets (FMs), DW velocities are significantly hindered by internal instabilities known as the Walker breakdown \cite{Schryer1974TheFields,Tatara2004TheoryTransfer,Hinzke2011DomainEffect} or by spin-wave (SW) emission analogous to the Cherenkov effect \cite{Akhmediev1995CherenkovFibers,Yan2013Spin-CherenkovCones}. AFMs however, can display relativistic DW kinematics limited only by the maximum group velocity of the medium, typically of the order of tens of km/s 
%\cite{Zvezdin1979DynamicsFerromagnets,Haldane1983NonlinearState,}
\cite{Gomonay2016HighTorques,Otxoa2020Walker-likeFields}.
% Several unique properties motivate the growing appeal for this class of magnetic materials, including their typical resonance frequencies in the THz regime, \textbf{[Cite]} robustness against external field perturbations \textbf{[Cite]} and absence of magnetic stray fields 
\newline\indent  
 The state of the art mechanism employed to manipulate the AFM order relies on the relativistic, current-induced N\'eel Spin Orbit Torque (NSOT) effect \cite{Zelezny2014RelativisticAntiferromagnets} shown to induce switching \cite{Wadley2016Spintronics:Antiferromagnet,Roy2016RobustFields,Bodnar2018WritingMagnetoresistance,Olejnik2018TerahertzMemory,Godinho2018ElectricallyAntiferromagnet} and DW motion \cite{Gomonay2016HighTorques,Otxoa2020GiantMetals,Janda2020Magneto-SeebeckCuMnAs,Otxoa2020Walker-likeFields}.
 %, arising in specific crystal structures and manifesting through staggered spin-orbit fields capable of inducing switching \cite{Wadley2016Spintronics:Antiferromagnet,Roy2016RobustFields} or DW motion \cite{Gomonay2016HighTorques,Otxoa2020Walker-likeFields}.
 % In specific crystal structures where neighbouring magnetic atoms in distinct sublattices form non-centrosymmetric local inversion partners, a current excitation gives rise to a non-equilibrium spin polarisation of alternating sign with respect to the AFM bipartite subsystem,
At the same time, laser pulse excitation displays the fastest and least dissipative control of the magnetic order observed so far \cite{Kimel2019WritingPulses}. Several studies addressed the question of light-matter interaction in AFMs \cite{Kimel2004Laser-inducedTmFeO3,Kampfrath2010CoherentWaves,Wienholdt2012THzFerrimagnets,Disa2020PolarizingField,Dannegger2021UltrafastEffect, Hedrich2021NanoscaleWalls,Grigorev2022OpticallyStrain,Weienhofer2023NeelTorques},
with THz laser pulse excitation shown recently capable of driving ultra-fast, non-linear dynamics in Mn$_2$Au via NSOT \cite{Behovits2023TerahertzMn2Au}.
However, the demonstration of all-optical AFM DW dynamics is yet to be achieved. 
% \textbf{Questions to self: Do we include THz excitation of spin-orbit torques in the category of laser-induced magnetisation dynamics? If so, what is the novelty in our case? optical frequencies compared to THz, IFE compared to NSOT.. Have AFMs DWs been displaced at THz or visible freqs. in any AFM? Are we 100 percent certain this statement is true: "to this day and to the best of our knowledge, the displace-
% ment of AFM DWs via the sole use of ultra-fast laser pulses
% has not been demonstrated yet." ? }
\newline\indent
A recent \textit{ab-initio} investigation concerning the symmetry and magnitude of the laser-induced magnetic response in the layered AFM Mn$_2$Au \cite{Freimuth2021Laser-inducedAntiferromagnets,Merte2023PhotocurrentsMn2Au} revealed the appearance of a substantial torque on the N\'eel order parameter $\bm{l}$ of tens of mT.
\begin{figure*}[!ht]
    \centering
    \includegraphics[width=1.0\linewidth]{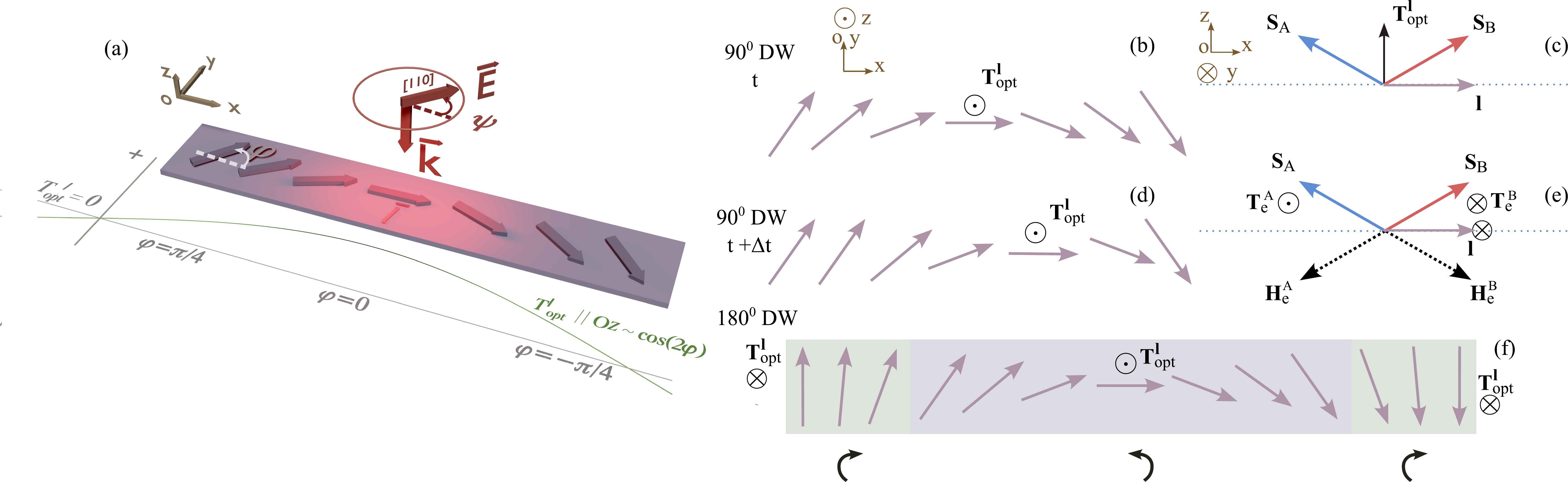}
    \caption{All-optical 90$^{\circ}$ DW dynamics in Mn$_2$Au. (a) A laser pulse in the optical spectrum propagating with $\bm{k}||Oz$, and with $\bm{E}$ polarised along the $[1 1 0]$ diagonal ($\psi=45^\circ$), will lead to an OOP torque determined by the local spin orientation and proportional to the laser intensity \cite{Freimuth2021Laser-inducedAntiferromagnets}. 
    %The angles $\varphi$ and $\psi$ are defined with respect to the $Ox$ axis.
    (b) At a time $t$, the optical torque gives rise to an OOP canting of the sublattices $\sm_\text{A}$, $\sm_\text{B}$ \textemdash seen in (c). 
    %Under the strong AFM coupling, 
    Large exchange fields $\bm{H}_\text{e}^\text{A}$, $\bm{H}_\text{e}^\text{B}$ torque the bipartite system such that the N\'eel vector rotates counter clockwise in the $Oxy$ plane, promoting the displacement of the $90^{\circ}$ DW in the subsequent time step $t+\Delta t$ (d-e). 180$^{\circ}$ DW motion under the same optical torque symmetry is not allowed. 
    %In this geometry (subplot f), $\bm{T}_{\text{opt}}^l$ changes sign twice, such that the boundaries (green shaded) and wall region (purple shaded) rotate in opposite directions, leading to expansion/contraction of the DW. In the 90$^{\circ}$ configuration, the optical torque preserves the same sign within the boundaries of the wall such that a preferential direction of motion can be set. 
    %In subplot (a) we preferred the arrow to bold vector representation for clarity.
    } 
    \label{fig:figure2}
\end{figure*}
Within the non-equilibrium Keldysh formalism \cite{Freimuth2016Laser-inducedFerromagnets,Freimuth2021Laser-inducedAntiferromagnets}, a linearly or circularly polarised laser pulse excitation at optical frequencies was shown to excite such a torque 
% a torque on the N\'eel vector parameter in the Mn$_2$Au crystal 
through the appearance of a staggered opto-magnetic field which couples to the Mn spins in the distinct AFM sublattices. The physical origin of the calculated effect is being attributed collectively to the Optical Spin Transfer Torque \cite{Nemec2012ExperimentalTorque,Ramsay2015OpticalSemiconductor} and the Inverse Faraday Effect \cite{Kimel2005UltrafastPulses,Quessab2018Helicity-dependentFilms}. Recently, this optical torque mechanism was predicted capable of inducing coherent $90^\circ$ switching of the AFM order parameter on the ultra-fast time-scale \cite{Ross2023AntiferromagneticTimescales}. 

% Inspired by the potential technological realisation of an all-optical, AFM DW spintronic device,
Here, we combine the \textit{ab-initio} results in Ref. \cite{Freimuth2021Laser-inducedAntiferromagnets} with atomistic spin dynamics (ASD) simulations \cite{Evans2014AtomisticNanomaterials}, in an investigation of optically-driven DW kinematics in Mn$_2$Au.
Depending on the relative orientation between the N\'eel vector $\bm{l}$ and the electrical field component $\bm{E}$ of the applied laser pulse, the symmetry and magnitude of the induced optical torque can be established.
%The magnetic configuration of the Mn$_2$Au crystal reveals four possible optical torque geometries which can be accessed via a linearly  polarised laser pulse. 
%The main result of our work demonstrates how 
We demonstrate that one of these laser-induced torques predicted in \cite{Freimuth2021Laser-inducedAntiferromagnets} may  drive 90$^{\circ}$ DW dynamics. Interestingly, none of them 
%allowed optical torque symmetries 
can be used to drive 180$^{\circ}$ DWs. Our numerical study is complemented by a theoretical,  two-sublattice  $\sigma$-model \cite{Rama-Eiroa2022InertialAntiferromagnets}, adapted for the present case.
%previously discussed in the context of current-induced 180$^{\circ}$ DW dynamics in Mn$_2$Au \cite{Rama-Eiroa2022InertialAntiferromagnets}.
\noindent
%\section{Numerical model}
% \begin{figure}
%     \centering
%     \includegraphics[width=0.5\textwidth]{Figures/Figure_1.pdf}
%     \caption{\emph{y axis needs to go other direction to have right hand rule coordinate. This will also change phi $x_0$ to pi/4, and phi $x_f$ to 3pi/4, which matches simulation and plot of figure 4.-JLR} Optical excitation of a one dimensional Mn$_2$Au system. A single laser pulse propagating with $\bm{k}||Oz$ and with the electric field component $\bm{E}$ polarised along the $Oxy$ diagonal, will lead to an out of plane optical torque proportional to the local spin orientation and the intensity of the laser pulse. \cite{Freimuth2021Laser-inducedAntiferromagnets} In a $90^0$ DW geometry, the spins at the boundaries will experience no torque while the spin corresponding to $\varphi=\pi/2$ will experience a maximum torque. The azimuthal angle $\varphi$  is defined with respect to the $Oy$ axis. }
%     \label{fig:Figure_1}
% \end{figure}
\newline\indent
The extended Heisenberg Hamiltonian in our Mn$_2$Au system is 
\cite{Roy2016RobustFields,Otxoa2020Walker-likeFields,Rama-Eiroa2022InertialAntiferromagnets}:  
\begin{align}
\mathcal{H}=&-\sum_{\langle i,j\rangle}J_{ij}\sm_i\cdot\sm_j-K_{2\perp}\sum_i(\sm_i\cdot\bm{z})^2-\frac{K_{4\perp}}{2}
\sum_i(\sm_i\cdot\bm{z})^4 \nonumber\\
%-K_{2||}\sum_i(\sm_i\cdot\bm{y})^2
&-\frac{K_{4||}}{2}\sum_{i}\left[(\sm_i\cdot\bm{u}_1)^4+(\sm_i\cdot\bm{u}_2)^4\right]-\mu_0\mu_\text{s}\sum_i\sm_i\cdot\bm{H}^i_{\text{opt}.} \label{Hamiltonian}
\end{align}
The adimensional $\sm_{i,j}$ unit vectors denote the orientation of the local spin magnetic moments at sites $i$ and $j$. The first energy term characterises the nearest-neighbours exchange interactions between $\sm_i$ and $\sm_j$. $J_{ij}=(J_1,J_2,J_3)$ collectively denotes three interactions present within the unit cell, two of AFM kind ($J_1$, $J_2$) and a third FM-like exchange ($J_3$). 
%, all previously parametrised by Khmelevskyi and Mohn \cite{Khmelevskyi2008LayeredAu}.
The Hamiltonian also contains three magnetocrystalline anisotropy terms  which  stabilise an in-plane $90^{\circ}$ DW configuration: one easy-plane contribution $K_{2\perp}$ and two tetragonal terms $K_{4\perp}$ and $K_{4||}$.
% Although not shown here, the nucleation of an $180^{\circ}$ DW would require the presence of a second-order, uni-axial term along $Oy$ and the removal of the $K_{4||}$ term.
The unit vectors $\bm{u}_1$,$\bm{u}_2$ denote the in-plane, diagonal directions $[110]$ and $[1\bar{1}0]$, respectively. The last energy term describes the Zeeman interaction with the staggered optical field $\bm{H}_{\text{opt}}^i$, where $\mu_\text{s}$ is the Mn atomic magnetic moment and $\mu_0$ is the vacuum permeability \cite{supplemental}.
\newline\indent
%Following the theoretical description of R. Rama-Eiroa \textit{et al.}, \cite{Rama-Eiroa2022InertialAntiferromagnets} we analytically characterise the Mn$_2$Au system in terms of a two-sublattice model which takes into account the pair magnetisation vectors ($\sm_A$, $\sm_B$) of the embedded planes in the four-layered unit cell crystal. In this approach... 

%We define a N\'eel order parameter $\bm{l}=(\sm_\text{B}-\sm_\text{A})/2$ and an effective magnetisation vector $\bm{n}=(\sm_\text{A}+\sm_\text{B})/2$, where $\sm_\text{A}$, $\sm_\text{B}$ stand for the pair magnetisation vectors of the embedded planes
%in the four-layered unit cell crystal. 

A linearly polarised pulse, with the electric field $\bm{E}$ parallel to the $[110]$ direction, leads to an optical torque $\bm{T}^{\bm{l}}_{\text{opt}}$ canting the N\'eel vector $\bm{l}$ in the $Oz$ direction as described by tensor 24 in Table 1 of Ref \cite{Freimuth2021Laser-inducedAntiferromagnets}. According to spatial symmetry arguments, the induced optical torque 
%at a given pulse intensity 
depends on the local orientation of the N\'eel vector and the $\bm{E}$-field as:
\begin{equation}
\bm{T}^{\bm{l}}_{\text{opt}}=A\cos(2\varphi)\sin(2\psi)\bm{\hat{z}},
\label{Neel_vector_torque}
\end{equation}
where $\varphi$ and $\psi$ measure the azimuthal deviation of $\bm{l}$ and $\bm{E}$ with respect $Ox$ (Figure \ref{fig:figure2}a). To maximize the optical torque we consider $\psi=45^\circ$. The amplitude $A$ expressed in units of T, scales linearly with the laser intensity $I$ and is adjusted by the quasi-particle broadening $\Gamma$, the latter being a measure of the disorder and the lifetime of the electronic states within the Keldysh formalism \cite{Freimuth2016Laser-inducedFerromagnets}. For a typical laser intensity of $I=10$ GW/cm$^2$, photon energy of $\nu=1.55$ eV and $\Gamma=25$ meV, $A$ evaluates to approximately $0.346$ T. This torque corresponds to a staggered opto-magnetic field $\bm{H}_{\text{opt}}^i$
%, which we couple to the local spin $\sm_i$ 
via the relationship:
\begin{equation}
\bm{H}^{i}_{\text{opt}}=\frac{1}{\mu_0}\frac{\bm{T}^{\bm{l}}_{\text{opt}}\times\sm_i}{2},
\label{optical_field}
\end{equation}
%where the $1/2$ factor ensures the optical response obtained with respect to the N\'eel vector is normalised to a two sublattice magnetisation system. 
\newline\indent
For the investigation of DW dynamics we numerically integrate the  Landau-Lifshitz-Gilbert  equation
%\begin{equation}
%\frac{d\sm_i}{dt}=-\frac{\gamma}{1+\lambda^2}\sm_i\times\left[\bm{H}_{\text{eff}}^i+\lambda\left(\sm_i\times\bm{H}_{\text{eff}}^i\right)\right],
%\end{equation}
%where $\gamma$ is the electron gyromagnetic ratio and $\lambda$ is an adimensional, phenomenological damping parameter. 
with the effective field  acting at each atomic site $i$  given by: $\bm{H}_{\text{eff}}^i=-\frac{1}{\mu_0\mu_\text{s}}\frac{\delta\mathcal H}{\delta\sm_i}$ and damping parameter $\lambda$ \cite{Evans2014AtomisticNanomaterials}.
%No coupling to a thermal bath is considered, the dynamics being investigated at $0~K$. 
The simulated system is $7~     \mu$m in length, four Mn wide and four Mn tall, with periodic boundary conditions along $Oy$ and 
 open boundary conditions along $Ox$ and $Oz$. 
\newline\indent
In Figure \ref{fig:figure2}(b) we schematically represent the top view of the track at a time $t$ during the laser excitation. Focusing on the atomic site where $\bm{l}$ lies parallel to $Ox$ ($\varphi=0$), we appreciate the appearance of an optical torque canting the sublattices $\sm_\text{A}$, $\sm_\text{B}$ towards $Oz$, as graphically detailed in (c). Under the strong AFM coupling, local exchange fields $\bm{H}_\text{e}^\text{A}$, $\bm{H}_\text{e}^\text{B}$ torque the bipartite system such that the N\'eel vector rotates counter clockwise in the $Oxy$ plane, driving the 90$^{\circ}$ DW in the subsequent time step $t+\Delta t$ towards $+Ox$, as seen in (d) and (e). The DW motion direction depends on the product of DW chirality and the torque direction \cite{supplemental}. Furthermore, a $90^\circ$ rotation of the $\bm{E}$-vector to the $[1\bar 10]$ ($\psi=-45^\circ$) direction changes the sign of the optical torque and consequently the direction of motion. In turn, a $180^\circ$ rotation to $[\bar 1 \bar 10]$ ($\psi=-135^\circ$) preserves the positive sign of the torque and the motion direction towards $+Ox$ \cite{supplemental}. 
\newline\indent
The $\cos(2\varphi)$ spatial variation of $\bm{T}_{\text{opt}}^l$ hinders the dynamics of an 180$^{\circ}$ DW. The key factor which enables the $90^\circ$ displacement is the optical torque acting along the same direction (parallel or antiparallel to $Oz$) at all the atomic sites within the boundaries of the wall such that a preferential DW sense of motion can be established. This resembles the NSOT-driven dynamics  in the 180$^{\circ}$ DW geometry \cite{Gomonay2016HighTorques,Otxoa2020Walker-likeFields}: the NSOT symmetry is $\sin(\varphi)$  which leads to zero torque at the boundaries of the 180$^{\circ}$ configuration and preserves the same sign torque across the transition region between domains. 
Contrarily, $\bm{T}_{\text{opt}}^l$ changes sign twice in an 180$^{\circ}$ geometry such that the boundaries (green shaded) and the wall region (purple shaded) tend to rotate in opposite directions (clockwise and counter-clockwise, respectively) leading to an expansion/contraction of the DW texture but no overall displacement, see Figure \ref{fig:figure2}(f). A second out-of-plane (OOP) optical torque labeled tensor 4 in Ref. \cite{Freimuth2021Laser-inducedAntiferromagnets} is predicted to arise for light polarised either along $Ox$ or $Oy$, given an in-plane N\'eel vector. This term is characterised by the $\sin(2\varphi)/2$ angular variation, which changes sign across the 90$^{\circ}$ wall and thus cannot move it. 
\begin{figure}[!ht]
    \centering    \includegraphics[width=0.7\linewidth]{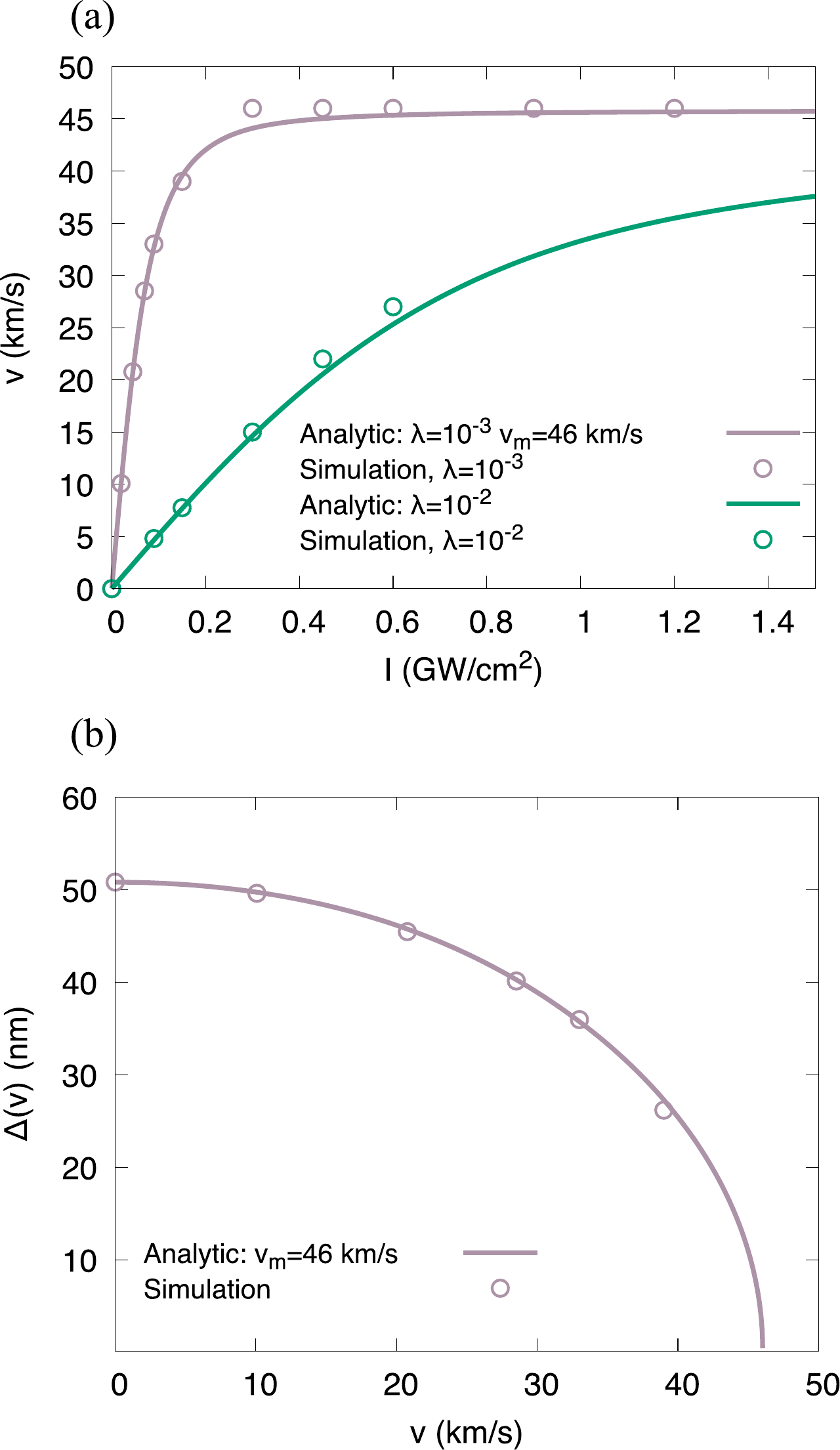}
    \caption{Numerical results and SG predictions of steady-state 90$^{\circ}$ DW dynamics: (a) Velocity saturation with the laser intensity for $\lambda=10^{-2}$ and $\lambda=10^{-3}$. (b) Contraction of the DW width as a function of the propagation velocity. Theory and calculations yield the same wall width at rest (up to two decimals): $\Delta_0=50.82$ nm. The numerical results overestimate the maximum velocity to a value of $46$ km/s compared to the analytical prediction of $43.39$ km/s. For better fitting, we adjust $v_\text{m}$ to the former value in subplots (a) and (b).   %    As discussed in the main text, this difference may arise due to SW excitation near the DW tail which broadens the wall and asymmetrically modifies its shape, deviating from the rigid profile assumed in the SG model.
    }
    % \textbf{Note: subplot b) is momentarily represented as a combined plot containing numerical points obtained with $\lambda=1e-3$ and $\lambda=1e-2$.}}
    \label{fig:Figure_3}
\end{figure}
%Moreover, in our $\bm{E}$-field geometry this contribution is shown to be zero. 
The two other in-plane torques, labeled $3$ and $9$ in Ref. \cite{Freimuth2021Laser-inducedAntiferromagnets},  arise for $\bm{E}$ polarised in the $Oxz$ or $Oyz$ planes. However, these torques cannot efficiently move DWs due to  the large $K_{2\perp}$ anisotropy which competes against  the canting of the individual sublattices in the $Oxy$ plane.
\newline\indent
A well known behaviour of AFM solitons, mainly discussed previously for 180$^{\circ}$ DWs \cite{Gomonay2016HighTorques,Otxoa2020Walker-likeFields}, the laser-induced dynamics of our 90$^{\circ}$ wall  display special relativity signatures manifesting through wall-width contraction as the propagation velocity approaches the "magnonic barrier" \textemdash see Figure \ref{fig:Figure_3}. 
%This is revealed in the ASD simulations by applying a constant, laser-induced torque of  increasing strengths which allows us to access a steady-state DW motion regime. 
%The numerical results compare well with analytical predictions obtained within a two-sublattice, $\sigma$ non-linear model. 
%which maps the AFM LLG equations of motion expressed for the $\bm{l}$ and $\bm{n}$ vectors onto a Lorentz-invariant, sine-Gordon (SG) like equation \cite{Rama-Eiroa2022InertialAntiferromagnets}.  
Similarly to NSOT-driven 180$^{\circ}$ DW dynamics \cite{Rama-Eiroa2022InertialAntiferromagnets},  the kinematics of the 90$^\circ$ DW is reduced to a Lorentz-invariant, sine-Gordon (SG) like equation
%second-order, differential equation 
in terms of the azimuthal angle $\varphi$ \cite{supplemental}:
\begin{equation}
\partial_x^2\varphi-\frac{1}{v_\text{m}^2}\ddot{\varphi}-\frac{1}{4\Delta_0^2}\sin4\varphi-h_{\text{opt}}\sin(2\varphi)-\eta\dot\varphi=0.
\label{SG_equation}
\end{equation}
\begin{figure*}[!ht]
    \centering
\includegraphics[width=1.0\textwidth]{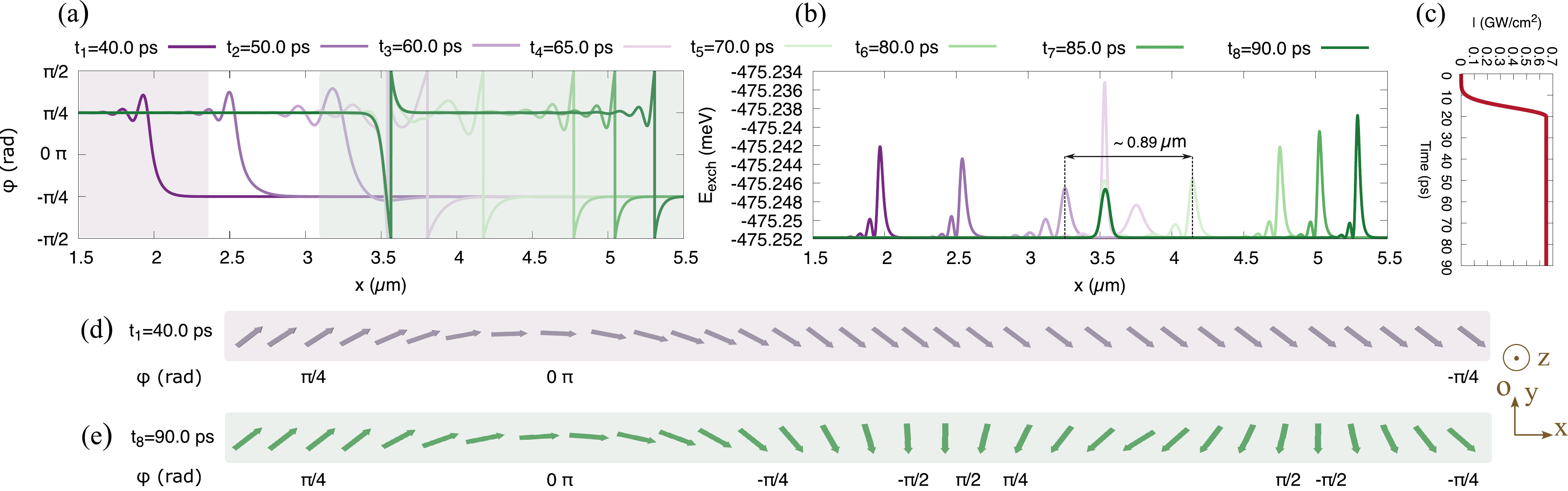}
     \caption{Proliferation event in the supermagnonic regime. (a) $\varphi$($x$,$t$) variation along the track during the 90$^{\circ}$ DW displacement.
     Time snapshots $t_1$, $t_2$ and $t_3$ showcase a gradual wall width broadening and a corresponding decrease in the exchange energy penalty $E_{\text{exch}}$ (b) under the influence of the lagging SW mode. A subsequent rebound contraction and increase of $E_{\text{exch}}$ takes place due to the competition between the exchange and anisotropy energies, leading to oscillatory patterns at the front of the propagating wall, which result into a proliferation event \textemdash seen at $t_4$ and $t_5$. The initially moving DW morphs into a static magnetic pattern pinned around the $x=3.5$ \textmu m mark, while the surplus relativistic energy is invested towards pushing ahead the novel spin structure. The energy transfer from the parent to the child magnetic texture takes place with an average velocity of $\approx 89$ km/s, extracted by comparing snapshots $t_3$ and $t_5$. Subplots (a) and (b) share the time legend. (c) Optical excitation protocol with peak laser intensity of $I=0.65$ GW/cm$^2$.  (d-e) Schematics of the N\'eel vector configuration corresponding to the purple and green shaded regions in (a), at $t_1=40$ ps and $t_8=90$ ps respectively.}
    \label{fig:Figure_4}
\end{figure*}
%T%he first two terms express the classical wave equation, where 
Here $v_\text{m}$ is the effective "speed of light" or the magnonic limit, encoded by the exchange interactions 
%present within the unit cell 
and evaluated in our case to $v_\text{m}=4\sqrt{a|J_1|}/(4\hbar)=43.39$ km/s, where  $a=8a_0^2(J_3+|J_1|/2)$ and $a_0$ is the spin lattice constant along $Ox$ and $Oy$
%as described for Mn$_2$Au in 
\cite{Rama-Eiroa2022InertialAntiferromagnets,supplemental}. 
%The effective exchange parameters $a$, $A$ are defined as: $a=8a_0^2(J_3+|J_1|/2)$ and $A=16|J_1|$, respectively, where $a_0$ is the spin lattice constant along $Ox$ and $Oy$. 
%The third term of equation \eqref{SG_equation} stands for a field-like magneto-crystalline anisotropy arising from the $K_{4||}$ contribution required to describe a 90$^{\circ}$ DW configuration and intrinsically present within the definition of the domain wall width at rest 
Furthermore, $\Delta_0=\sqrt{a/(8K_{4||})}=50.8$ nm is the DW width at rest and the reduced optical field $h_{\text{opt}}$ is defined as $h_{\text{opt}}=8\gamma\hbar H_{\text{opt}}/a$.
%, where the magnitude $H_{\text{opt}}$ is obtained from equation \eqref{optical_field}.
The last term represents the phenomenological damping field 
%whose amplitude is described in terms of the constant 
$\eta=8\lambda\hbar/a$. 
In a steady-state motion regime, when the optical excitation counterbalances the dissipation, Eq. \eqref{SG_equation} admits the following solution for the 90$^{\circ}$ DW profile: $\varphi(x,t)=\arctan[\exp((x-vt)/\Delta(v))]$. The equilibrium between the damping and the optical excitation ensures the Lorentz invariance of the dynamical equation \eqref{SG_equation} and yields the velocity-dependent wall width $\Delta(v)=
%\Delta_0\beta(v)
\Delta_0\sqrt{1-\frac{v^2}{v_\text{m}^2}}$. Here 
%$\beta$ stands for the reciprocal Lorentz factor and 
$v$ is the steady-state velocity 
%Equating the last two terms in \eqref{SG_equation}, we can obtain the following analytical expression for 
$v=\frac{v_\text{m}}{\sqrt{1+(v_\text{m}/v_0)^2}}$,
where $v_0=2h_{\text{opt}}\Delta_0/\eta$. 
\newline\indent
In Figure \ref{fig:Figure_3}, we analyse the numerical wall width and velocity in the steady-state regime at different laser pulse intensities and compare them with the analytic expressions presented earlier. The  simulations compare well with the developed theory \cite{supplemental}, below the predicted magnonic barrier. 
%\sout{ As previously discussed by Yang \textit{et al.} \cite{Yang2019AtomicLimit}, this SW excitation will lead to a broadening of the DW's width and will augment its propagation velocity beyond the expected result described by the continuum SG theory}. 
%According to the work of R.M. Otxoa \textit{et al.,} \cite{Yang2019AtomicLimit} 
\newline\indent
Similar to AFM DW driven by NSOT \cite{Otxoa2020Walker-likeFields}, optically-excited DW dynamics  near the magnonic barrier  enters a highly non-linear regime,
%akin to the Walker-breakdown behaviour in ferromagnets, 
leading to the proliferation of additional magnetic textures, breaking of the Lorentz invariance and supermagnonic motion, see Figure \ref{fig:Figure_4}.
%a similar effect can occur in our optical excitation protocol for a sufficiently large pulse intensity.
Since the Zeeman energy contribution always remains zero in our case, as the geometry in Eq. \eqref{optical_field} maintains $\bm{H}_{\text{opt}}^i$ always perpendicular to the local spin direction $\sm_i$, the question is, therefore, where does the energy required for the nucleation of additional magnetisation textures arise from? In subplot (a) of Figure \ref{fig:Figure_4}, we represent the azimuthal angle $\varphi$ along the track during a $90^\circ$ wall displacement excited via a half-Gaussian, laser pulse of peak intensity $I=0.65$ GW/cm$^2$ (see laser profile in subplot c). Time-steps $t_1$, $t_2$, $t_3$ show the characteristic low-frequency SW, lagging behind and broadening the width of the 90$^{\circ}$ DW as it is pushed beyond the magnonic barrier. A gradual, corresponding decrease in the exchange energy is evidenced by subplot (b). Due to the competition between the anisotropy and exchange energies, this broadening cannot indefinitely take place, forcing a rebound process characterised by a contraction of the wall. A drastic reduction of the DW width can be observed between $t_3$ and $t_4$ along with a large increase in the exchange energy. This continuous deformation leads to the appearance of oscillatory patterns at the front of the propagating wall, which on a ps time-scale invest part of the relativistic energy carried by the DW towards the nucleation of an additional magnetisation texture, as evidenced at $t_4$ and $t_5$.
% On average, this relativistic energy is being transferred \textbf{Here maybe we could approximate the speed of the energy transfer and put in on the graph also-  it is on average around 100 km/s in between timeframes $t_4$ and $t_5$ - it travels 0.5 microns in 5 ps.} 
Unlike the 180$^{\circ}$ NSOT-driven DW case \cite{Otxoa2020Walker-likeFields}, the newly formed pattern does not follow the geometry of the parent texture as further evidenced by subplots (d) and (e).
%, but in identical fashion, the topological charge of the system before and after the nucleation event is preserved \Jackson{the idea of topological protection is new at this point in the paper. Maybe remove this comment, or move it to the end of this paragraph?}
Visible at $t_6$, $t_7$ and $t_8$, the initial spin structure morphs into a static magnetisation texture pinned around the $x=3.5$ \textmu m mark, while its surplus, relativistic energy is invested towards pushing ahead a novel spin structure. 
%Similar to the kinematics before proliferation, a low frequency SW  can be observed lagging behind the newly nucleated texture.
The pair spin textures are stable in a reasonable  timeframe achieved by atomistic simulations after the laser pulse is stopped and provided they do not overlap.
\newline\indent
The physical origin of the effect relies on the rapid transfer of relativistic energy from a propagating DW near the magnonic barrier into new stable magnetisation textures. Comparing snapshots $t_3$ and $t_5$, we estimate the energy transfer across the track from the parent to the child magnetic texture takes place with an average velocity of $89$ km/s ($0.89$ \textmu m travelled in 10 ps), largely exceeding the speed limit of a steady state soliton, also called the "effective speed of light". 
This breakdown of the Lorentz invariance  and soliton-like behavior 
occurs due to dynamical changes of the DW shape which result in the creation of new textures with repulsive interaction leading to acceleration.
Interestingly, these results can find similarities with dislocations dynamics. As shown in \cite{Stroh1962,FrankMerwe}, dislocations may enter a super-sonic regime surpassing the sonic barrier, the underlying reason for this being the generation of secondary kinks (topological defects) via the so-called "mother-daughter" kink nucleation. 
%\textcolor{blue}{unanswered questions: 1. Why is the exchange energy increasing after 90 ps? Does the proliferation process go on? 2. Why do we even see proliferation? Do you think citing previous works who saw but didn't explain this is sufficient to avoid painful referee questions about this?} 
%Understanding and tailoring this highly non-linear dynamics regime may push AFM DW based spintronics beyond the special relativity boundaries.
\newline\indent
In conclusion, our  simulations revealed the possibility to drive a $90^{\circ}$ AFM DW  under a special optical torque symmetry.  Interestingly this excitation protocol does not allow the motion of $180^{\circ}$ DWs. 
%This work serves as a proof of concept where the fundamental question of magneto-optics in AFM systems is approached in a simplified one-dimensional picture at $0~K$. 
%The high N\'eel temperature of Mn$_2$Au set around $1600~K$ as well as the reduced laser pulse intensities we require to displace the $90^0$ DW configuration motivate in a first approximation the  exclusion of thermal effects. 
Typical experiments of laser-induced DW dynamics in FMs make use of fluences in the range of a few mJ/cm$^2$ which amount to intensities normally in the $1$ and $10$ GW/cm$^2$ interval \cite{Quessab2018Helicity-dependentFilms,Parlak2018OpticallyDynamics}. In comparison, we predict 90$^{\circ}$ DW kinematics up to the supermagnonic limit ($v=46$ km/s) by single pulse excitation below $0.3$ GW/cm$^2$. 
%We estimate using a two-temperature model the effect of the heating dynamics in the range of laser parameters used here. 
The corresponding phonon heating is on the timescale below 1 ns and is estimated   \cite{supplemental} much lower than  Si or Al$_2$O$_3$ substrate melting temperature \cite{Behovits2023TerahertzMn2Au}  and the N\'eel temperature of Mn$_2$Au. 
\newline\indent
Manipulating highly non-linear dynamics in magnetic systems may play an important role in the development of future reservoir computing archetypes \cite{Kurenkov2020NeuromorphicSpintronics}. We envision thus a combined, opto-electronic experimental scheme, towards the realisation of a multiple-node reservoir. Driven by an optical input, fast and periodic nucleation events could be manipulated in an AFM DW fabric \cite{Bourianoff2018PotentialSkyrmions}, whereas based on the anisotropic magnetoresistive effect \cite{Bodnar2018WritingMagnetoresistance}, an output electrical signal may be correlated to changes in the magnetic texture, thus posing an interest for pattern recognition and prediction applications. 
% Similar to neuromorphic devices based on spin-torque oscillators \cite{Kamimaki2021ChaosCircuit,Imai2022Input-drivenOscillator}, by means of dynamic polarisation modulation and combined electrical tuning \cite{Misawa2016ApplicationsPulses,Qi2021ManipulationApplications,Perosa2023FemtosecondPulses},... could be relayed into a feedback circuit to generate nucleation processes while optical excitation drives fast and periodic 90$^{\circ}$ DW dynamics at relativistic speeds. 
Lastly, the diversity and natural abundance of antiferromagnetically ordered materials calls for an exploration of similar optical torque symmetries in other crystal structures, as well the natural extension towards the emergent landscape of altermagnets \cite{Smejkal2022EmergingAltermagnetism}.
 \newline\indent
This project has received funding from the European Union’s Horizon 2020 research and innovation programme under the Marie Skłodowska-Curie International Training Network COMRAD, grant agreement No 861300. The atomistic simulations were undertaken on the VIKING cluster, which is a high performance compute facility provided by the University of York. F.F. and Y.M. acknowledge the funding by the Deutsche Forschungsgemeinschaft 
(DFG, German Research Foundation) $-$ TRR 173/2 $-$ 268565370 (project A11) 
and Sino-German research project DISTOMAT (MO 1731/10-1). The work of O.C.-F. has been supported by DFG  via CRC/TRR 227, project ID 328545488 (Project MF).

% \section{Appendixes}


%\nocite{*}
% \cleardoublepage
% \phantomsection
\bibliography{References_byhand.bib}
% Produces the bibliography via BibTeX.

\end{document}

%\preprint{APS/123-QED}

\title{Efficient motion of 90$^{\circ}$ domain walls in Mn$_2$Au via pure optical torques
\\Supplementary information} % Force line breaks with \\

\author{Paul-Iulian Gavriloaea}
\affiliation{Instituto de Ciencia de Materiales de Madrid, CSIC, Cantoblanco, 28049 Madrid, Spain}
\author{Jackson L. Ross}
\affiliation{School of Physics, Engineering and Technology, University of York, YO10 5DD, York, United Kingdom}
\author{Frank Freimuth}
\affiliation{Institute of Physics, Johannes Gutenberg University Mainz, 55099 Mainz, Germany}
\affiliation{Peter Grünberg Institut and Institute for Advanced Simulation, Forschungszentrum Jülich and JARA, 52425 Jülich, Germany}
\author{Yuriy Mokrousov}
\affiliation{Peter Grünberg Institut and Institute for Advanced Simulation, Forschungszentrum Jülich and JARA, 52425 Jülich, Germany}
\affiliation{Institute of Physics, Johannes Gutenberg University Mainz, 55099 Mainz, Germany}
\author{Richard~F.~L.~Evans}
\affiliation{School of Physics, Engineering and Technology, University of York, YO10 5DD, York, United Kingdom}
\author{Roy Chantrell}
\affiliation{School of Physics, Engineering and Technology, University of York, YO10 5DD, York, United Kingdom}
\author{Oksana Chubykalo-Fesenko}
\affiliation{Instituto de Ciencia de Materiales de Madrid, CSIC, Cantoblanco, 28049 Madrid, Spain}
\author{Rubén M. Otxoa}
\affiliation{Hitachi Cambridge Laboratory, CB3 OHE Cambridge, United Kingdom}
\affiliation{Donostia International Physics Center, 20018 Donostia San Sebastian, Spain}

\maketitle

%\tableofcontents
\onecolumngrid
\newpage
\section*{I. Atomistic spin dynamics simulation parameters}

The unit-cell geometry used in our work and the material parameters correspond to the previous works in Refs. \cite{Roy2016RobustFields,Otxoa2020Walker-likeFields,Rama-Eiroa2022InertialAntiferromagnets}. Thus, the lattice constants in the tetragonal unit cell are set to $a_0=3.328$ \textup{~\AA} along $Ox$, $Oy$ and $c=8.539$ \textup{~\AA} on $Oz$ \cite{Khmelevskyi2008LayeredAu}. 
The atomistic exchange constants evaluate to $J_1=-5.4673\times 10 ^{-21}$ J/link, $J_2=-7.3451 \times 10^{-21}$ J/link and $J_3=1.5877 \times 10^{-21}$ J/link \cite{Khmelevskyi2008LayeredAu}. Further, the anisotropy constants are given by $K_{2\perp}=-1.303\times10^{-22}$ J/atom, $K_{4||}=1.8548 \times 10^{-25}$ J/atom and $K_{4\perp}=3.7096 \times 10^{-25}$ J/atom \cite{Shick2010Spin-orbitSpintronics}. Finally, the Mn atomic magnetic moment is equal to $\mu_\text{s}=3.73$ $\mu_{\text{B}}$ \cite{Freimuth2021Laser-inducedAntiferromagnets}. Two distinct damping parameters are considered, $\lambda=10^{-2}$ and $\lambda=10^{-3}$. 

\section*{II. Details on the derivation of the sine-Gordon equation in the $90^\circ$ domain-wall geometry including the optical torque contribution}
Eq. 5 in the manuscript is obtained by mapping the Landau-Lifshitz-Gilbert (LLG) equations of motion for the magnetisation vectors ($\bm{S}_\text{A}$, $\bm{S}_\text{B}$) of the embedded layers in the Mn$_2$Au crystal onto a Lorentz-invariant, sine-Gordon (SG) like equation. 
To achieve this, we shall first rewrite the extended Heisenberg Hamiltonian (Eq. 1 in the manuscript) in terms of the orthogonal unit vectors $\bm{l}=\frac{1}{2}(\bm{S}_\text{B}-\bm{S}_\text{A})$ and $\bm{n}=\frac{1}{2}(\bm{S}_\text{A}+\bm{S}_\text{B})$ in similar fashion to  Refs. \cite{PhysRevB.93.104408,Rama-Eiroa2022InertialAntiferromagnets} and based on the pioneering work of Refs.  \cite{PhysRevLett.74.1859,PhysRevB.51.15062}: 
\begin{equation}
\label{rewrittenHam}
\mathcal{H}(\bm{n},\bm{l})=\frac{1}{2}A\bm{n}^2 + \frac{1}{8}a(\partial_x\bm{l})^2+|K_{2\perp}|(\bm{l}\cdot\bm{z})^2-\frac{K_{4\perp}}{2}(\bm{l}\cdot\bm{z})^4-\frac{K_{4||}}{2}\left[(\bm{l}\cdot\bm{x})^4+(\bm{l}\cdot\bm{y})^4\right]+2\gamma\hbar \bm{l}\cdot\bm{H}_{\text{opt}},
\end{equation}
where the first two terms encapsulate the exchange energy contributions in a continuum limit, governed by the $A=16|J_1|$ and $a=8a_0^2(J_3+|J_1|/2)$ constants respectively. It is important to note the AFM $J_2$ interaction along the $c$ axis of the tetragonal unit cell is disregarded in this analytical treatment, as it was shown previously to have a negligible influence on the in-plane $180^\circ$ DW dynamics of a Mn$_2$Au thin film \cite{Rama-Eiroa2022InertialAntiferromagnets}. We maintain this assumption in the case of our in-plane $90^\circ$ wall dynamics.
Differently to the description of $180^\circ$ DWs, our Hamiltonian includes the $K_{4||}$ anisotropy contribution in addition to the $K_{4\perp}$ and $K_{2\perp}$ terms, while it disregards the second-order, easy-axis contribution labelled $K_{2||}$ in \cite{Rama-Eiroa2022InertialAntiferromagnets}.
For ease of calculation we align the $K_{4||}$ easy-axes in this analytical model along $Ox$ and $Oy$. In the Zeeman-like energy term of equation \eqref{rewrittenHam} we made use of the relationship $\mu_0\mu_s=2\gamma\hbar$.

Let us now introduce the LLG equations of motion for our two sublattice magnetisation system ($\bm{S}_\text{A}$, $\bm{S}_\text{B}$) which in the limit $\lambda\ll1$ can be written in a Landau-Lifshitz (LL) form:
\begin{align}
\label{sublatticeA}
\dot{\bm{S}}_\text{A}&=-\gamma\bm{S}_\text{A}\times\bm{H}_{\text{eff}}^\text{A}-\gamma\lambda\bm{S}_\text{A}\times(\bm{S}_\text{A}\times\bm{H}_{\text{eff}}^\text{A}),\\
\label{sublatticeB}
\dot{\bm{S}}_\text{B}&=-\gamma\bm{S}_\text{B}\times\bm{H}_{\text{eff}}^\text{B}-\gamma\lambda\bm{S}_\text{B}\times(\bm{S}_\text{B}\times\bm{H}_{\text{eff}}^\text{B}).
\end{align}  
Introducing two effective magnetic fields $\bm{H}_{\text{eff}}^{\bm{n},\bm{l}}=-\frac{1}{2\gamma\hbar}\frac{\delta\mathcal{H}(\bm{n},\bm{l})}{\delta(\bm{n},\bm{l})}$ defined with respect to the Hamiltonian in equation \eqref{rewrittenHam} and corresponding to the $\bm{n}$ and $\bm{l}$ vectors, it can be shown $\bm{H}_{\text{eff}}^{\text{A},\text{B}}$ in equations \eqref{sublatticeA}, \eqref{sublatticeB} satisfy the following relationships:
\begin{align}
\label{effFieldA}
\bm{H}_{\text{eff}}^{\text{A}}&=\bm{H}_{\text{eff}}^{\bm{n}}+\bm{H}_{\text{eff}}^{\bm{l}},\\
\label{effFieldB}
\bm{H}_{\text{eff}}^{\text{B}}&=\bm{H}_{\text{eff}}^{\bm{n}}-\bm{H}_{\text{eff}}^{\bm{l}}.
\end{align}
Based on equations \eqref{sublatticeA}, \eqref{sublatticeB}, \eqref{effFieldA} and \eqref{effFieldB} it is possible to reduce the dynamics of our two sublattice magnetisation system in the exchange limit $|\bm{n}|\ll|\bm{l}|$ to the form below \cite{Baryakhtar1980NonlinearAntiferromagnets,PhysRevB.93.104408,Rama-Eiroa2022InertialAntiferromagnets}:
\begin{align}
\label{master}
\dot{\bm{l}}&=\gamma\bm{H}_{\text{eff}}^{\bm{n}}\times\bm{l},\\
\label{slave}
\dot{\bm{n}}&=(\gamma\bm{H}_{\text{eff}}^{\bm{l}}-\lambda\dot{\bm{l}})\times\bm{l}.
\end{align}
Writing the explicit form of the $\bm{H}_{\text{eff}}^{\bm{n}}$ term in equation \eqref{master} it can be seen the $\bm{n}$ vector plays only a "slave variable" role. Hence, the system of equations \eqref{master}, \eqref{slave} can be rewritten solely in terms of the N\'eel order parameter $\bm{l}$ \cite{Baryakhtar1980NonlinearAntiferromagnets,PhysRevLett.74.1859,PhysRevB.51.15062}:
\begin{equation}
\label{dynamics_just_L}
\bm{l}\times\left[\partial_x^2\bm{l}-\frac{1}{v_\text{m}^2}\ddot{\bm{l}}-\frac{4}{a}\frac{\partial \mathcal{H}_{\text{ani}}}{\partial \bm{l}}-\bm{h}_{\text{opt}}-\eta\dot{\bm{l}}\right]=0,
\end{equation}
where we collectively represent the distinct anisotropy contributions via the umbrella term $\mathcal{H}_{\text{ani}}=|K_{2\perp}|(\bm{l}\cdot\bm{z})^2-\frac{K_{4\perp}}{2}(\bm{l}\cdot\bm{z})^4-\frac{K_{4||}}{2}\left[(\bm{l}\cdot\bm{x})^4+(\bm{l}\cdot\bm{y})^4\right]$. The constant $v_\text{m}$ denotes the effective "speed of light" of the medium defined as $v_\text{m}=\sqrt{aA}/(4\hbar)$, the term $\bm{h}_{\text{opt}}=\frac{8\gamma\hbar}{a}\bm{H}_{\text{opt}}$ represents the reduced opto-magnetic field  while the constant $\eta=\frac{8\lambda\hbar}{a}$ encodes the strength of the dissipations.

To simplify the description of our in-plane, N\'eel vector dynamics let us introduce a spherical coordinate system which relates to the cartesian space in the following manner:
\begin{align}
\bm{l}\equiv \bm{u}_r&=\cos\varphi\bm{u}_x+\sin\varphi\bm{u}_y,\\
\bm{u}_{\varphi}&=-\sin\varphi\bm{u}_x+\cos\varphi\bm{u}_y,\\
\bm{u}_{\theta}&=-\bm{u}_z.
\end{align}
Finally, in this new coordinate system equation \eqref{dynamics_just_L} reduces to the familiar SG-like equation below:
\begin{equation}
\partial_x^2\varphi-\frac{1}{v_\text{m}^2}\ddot{\varphi}-\frac{1}{4\Delta_0^2}\sin4\varphi-h_{\text{opt}}\sin(2\varphi)-\eta\dot\varphi=0.
\end{equation}
We underline the K$_{4||}$ anisotropy term in the main text is defined with respect to the diagonal directions $\langle110 \rangle$ and $\langle 1\bar{1}0 \rangle$. For ease of calculation in the analytical model presented here, we perform a $45^\circ$ in-plane rotation such that the easy-axes point along the $Ox$ and $Oy$ cartesian directions instead. For this reason, the angular variation of the optical torque acting on the N\'eel vector $\bm{l}$ also needs to be adjusted here to 
$\propto\sin(2\varphi)$.
In a more clear picture, $\bm{h}_{\text{opt}}$ is thus described by:
\begin{equation}
\bm{h}_{\text{opt}}=\frac{8\gamma\hbar}{a}\bm{H}_{\text{opt}}=\frac{8\gamma\hbar}{a}\bm{T}^{\bm{l}}_{\text{opt}}\times\bm{l}=-\frac{8\gamma\hbar A\sin(2\varphi)}{a}\bm{u}_{\theta}\times\bm{u}_{r}=-h_{\text{opt}}\bm{u}_{\phi}.
\end{equation}

\section*{III. Spin texture chirality analysis in optically driven 90\text{$^\circ$} wall geometries}
\begin{suppfigure}[!ht]
    \centering
\includegraphics[width=1.0\textwidth]{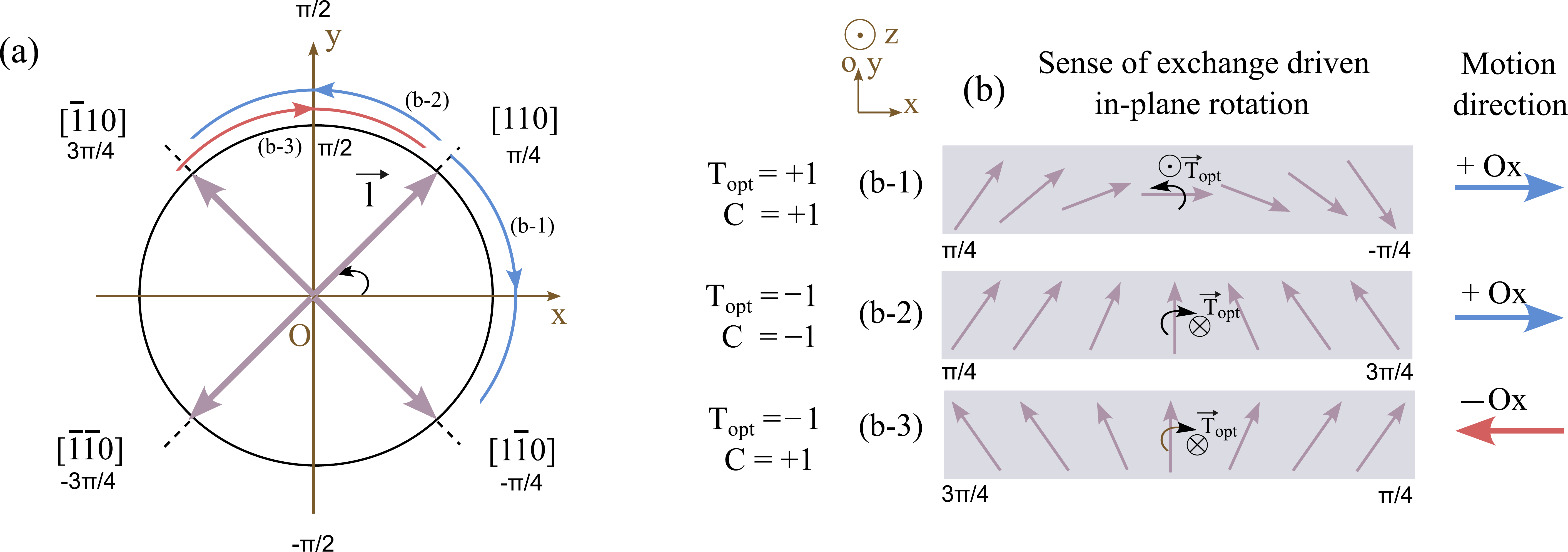}
     \caption{(a) Mapping the in-plane N\'eel vector orientation in the magnetic track along the unit circle. We display the four possible 90$^\circ$ DW boundaries corresponding to the $K_{4||}$ easy directions. Taking as reference the N\'eel vector parallel to the [110] diagonal, we can obtain two DW geometries with right and left hand side chirality, labeled (b-1) and (b-2) respectively. The corresponding spin configurations are displayed in subplot (b). ASD spin dynamics simulations reveal these two DWs displace in the same direction $+Ox$ because the handness of the in-plane N\'eel vector rotation opposes the chirality of the wall in both cases. Distinctly, the (b-3) wall geometry is driven towards $-Ox$ because the in plane spin reorientation matches the chirality of the wall.}
    \label{fig:figure_S1}
\end{suppfigure}
Mapping the spin configuration along the unit circle, one can trace eight possible 90$^\circ$ DW geometries in our bi-axial AFM. ASD simulations reveal four of these configurations will displace along $+Ox$ while the other four in the opposite $-Ox$ direction. The direction of motion depends on the sign of the optical torque relative to the chirality of the wall. We assign $C=+1$ for right-hand side DW chirality (clock-wise spin rotation) and $C=-1$ for left-hand side DW chirality (counter clock-wise spin rotation). Since the optical torque maintains its sign across the DW configuration, we can distinguish two situations, positive and negative optical torque $T_{\text{opt}}=\pm 1$. If the product $T_{\text{opt}}C$ is positive, the DW will displace towards $+Ox$, otherwise the motion will take place in the opposite direction. In Figure S\ref{fig:figure_S1} we exemplify this behaviour for three distinct DW configurations. Taking as a reference the N\'eel vector parallel to the $[110]$ direction in subplot (a), we construct two 90$^\circ$ DW configurations following a right or left-hand side rotation along the unit circle. The corresponding spin arrangements labeled (b-1) and (b-2) can be observed in subplot (b) of Figure S\ref{fig:figure_S1}. Both $T_{\text{opt}}$ and $C$ change their sign in between the two geometries, hence the product $T_{\text{opt}}C$ remains positive and the DW displacement direction is preserved. This can be understood in light of the strong in-plane AFM torque excited in the two-step driving mechanism discussed in Figure 1 of the manuscript. Depending on the sign of $T_{\text{opt}}$, the resulting in-plane rotation of the N\'eel vector will take place in opposite directions. $T_{\text{opt}}=+1$ gives rise to counter clock-wise rotation, while $T_{\text{opt}}=-1$ promotes clock-wise rotation. To maintain the DW direction of motion, the handness of this in-plane rotation must oppose the chirality of the wall. Therefore the $T_{\text{opt}}$ change of sign in between configurations (b-1) and (b-2) matches this requirement and thus preserves the direction of motion towards $+Ox$.
\newline\indent
The situation changes if we compare the (b-2) and (b-3) 90$^\circ$ DW profiles. Here the $T_{\text{opt}}C$ product changes sign which promotes displacement in opposite directions as confirmed by our ASD simulations. Taking as reference any other 
90$^\circ$ DW configuration along the unit circle in Figure S\ref{fig:figure_S1}(a), we observe the same behaviour. 

\section*{IV. Rotating the electric field of the linearly polarised laser} According to the symmetry analysis in Ref. \cite{Freimuth2021Laser-inducedAntiferromagnets} and equation 2 in the manuscript, we can account for a rotation of the $\bm{E}$-field vector with an angle $\psi$ with respect to $Ox$.
Hence, rotating the laser polarisation vector from the $[110]$ ($\psi=45^\circ$) to the $[1\bar 10]$ ($\psi=-45^\circ$) direction would change the sign of the optical torque. Taking as reference the spin configuration (b-1) in Fig. S1, the direction of motion of the DW changes when the sign of the optical torque is reversed. As discussed in section III of the supplemental material, this comes as a result of reversing the sign of the $T_{\text{opt}}C$ product. In contrast, a $180^\circ$ rotation of the laser polarisation from $[110]$ ($\psi=45^\circ$) to the $[\bar 1\bar 10]$ ($\psi=-135^\circ$) direction preserves the positive sign of the torque and the direction of motion towards $+Ox$.

\section*{V. Two-temperature model of laser induced heating}
We consider the physical effects from heating caused by the pulse durations and laser intensities used in our simulations on the Mn$_2$Au sample and substrate using the two-temperature model (TTM) described by Eqs. \eqref{TTM}, \eqref{TTM2}. Laser energy is added using the source term $S(t)$. Constants are from \cite{Otxoa2020GiantMetals} and are consistent with experimental results in \cite{Behovits2023TerahertzMn2Au,Grigorev2022OpticallyStrain}. $C_\text{e}=1 \times 10^3 $ J/K$^2$/m$^3$ is the electron heat capacity, $C_\text{p}=6.934 \times 10^6$ J/K/m$^3$ is the phonon heat capacity, $G=2.5 \times 10^{17}$ J/K/m$^3$ the electron-phonon coupling constant, and $\tau_\text{e}$ the substrate cooling factor. For electron temperature $T_\text{e}$, phonon temperature $T_\text{p}$, and substrate temperature $T_\text{s}$, the differential equation without diffusion is:
%\textcolor{orange}{Paul: I would modify Fig. S2 such that the intensity takes values on the right hand side y axis, y2.}
\begin{equation}
\label{TTM}
    C_\text{e} \frac{\text{d}T_\text{e}}{\text{d}t}=G(T_\text{p}-T_\text{e})+S(t)
\end{equation}
\begin{equation}
    C_\text{p} \frac{\text{d}T_p}{\text{d}t}=G(T_\text{e}-T_\text{p})+\frac{(T_\text{s}-T_\text{p})}{\tau_\text{e}}
    \label{TTM2}
\end{equation}
\begin{suppfigure}[!h]
    \centering
    \includegraphics{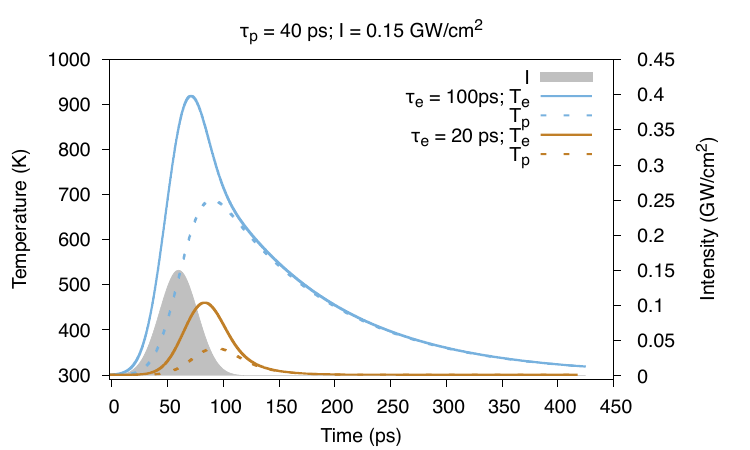}
    \caption{Two temperature model of Gaussian laser-pulse heating for peak intensity of $I=0.15$ GW/cm$^2$ and full width at half maximum of $40$ ps. We consider two distinct cooling rates of $20$ ps and $100$ ps. }
    \label{fig:TTM}
\end{suppfigure}

The relationship between laser intensity and laser power is described in detail in \cite{Jiang2005}. For simplicity we approximate a depth-independent absorption.

\begin{equation}
    S(t) = F\frac{2}{\delta \; t_\text{p}\sqrt{\pi /\text{ln}2}} \; \text{exp}\left[ -4 \text{ln}\;2 \left(\frac{t}{t_\text{p}}\right) \right],
\end{equation}
where,  $F$ is the fluence and $\delta$ the film thickness for which we assume 50 nm as in \cite{Behovits2023TerahertzMn2Au}. Since according to Fig. 2 of the manuscript, fast DW kinematics in the range of tens of km/s can be achieved in a wide laser intensity range, we make our estimations considering a Gaussian laser pulse with peak intensity $0.15$ GW/cm$^2$ and full width at half maximum of $40$ ps. The cooling rate which influences the peak electron and phonon temperatures depends largely on the substrate, for which here we consider two reasonably distinct values of $20$ and $100$ ps. According to the calculations presented in Fig. S2, the maximum electron/phonon heating can be contained only to a couple of hundreds of K above room temperature which is one order of magnitude lower compared to a Si or Al$_2$O$_3$ substrate melting temperature \textemdash the latter was recently used in Ref. \cite{Behovits2023TerahertzMn2Au} \textemdash and the N\'eel temperature of Mn$_2$Au.

% \cleardoublepage
% \phantomsection

\bibliography{References_supplemental.bib}

% Produces the bibliography via BibTeX.